\documentclass[aps,twocolumn,eqsecnum,prb]{revtex4}
\usepackage[T1]{fontenc}
\usepackage{amsmath,amsbsy,amssymb,graphicx}

\def\beginABC{\begin{subequations}}
\def\endABC{\end{subequations}}
\begin{document}

\title{{\Large Metallic Graphene Nanodisks}}
\author{Motohiko Ezawa}
\affiliation{{}Department of Physics, University of Tokyo, Hongo 7-3-1, 113-0033, Japan }

\begin{abstract}
We explore the electronic properties of finite-length graphene nanoribbons
as well as graphene nanodisks with various sizes and shapes in quest of
metallic ones. For this purpose it is sufficient to search zero-energy
states. We find that there exist no zero-energy states in finite-length
zigzag nanoribbons though all infinite-length zigzag nanoribbons have
zero-energy states. The occurrence of zero-energy states is surprisingly
rare. Among typical nanodisks, only trigonal zigzag nanodisks have
degenerate zero-energy states and show metallic ferromagnetism, where the
degeneracy can be controlled arbitrarily by designing the size. A remarkable
property is that the relaxation time is quite large in spite of its small
size in trigonal zigzag nanodisks.
\end{abstract}

\maketitle

%\date{}

\address{{\normalsize Department of Physics, University of Tokyo, Hongo
7-3-1, 113-0033, Japan }}

\section{Introduction}

Graphene\cite{GraphExA,GraphExB,GraphExC}, a single atomic layer of
graphite, has invoked a great interest in the electronic properties of
graphene-related materials. In particular, graphene nanoribbons\cite%
{Fujita,EzawaPRB,Brey,Rojas,Son,Barone,Kim,Avouris,Xu} have attracted much
attention due to a rich variety of band gaps, from metals to wide-gap
semiconductors. They are manufactured by patterning based on nanoelectronic
lithography methods\cite{Berger,Kim,Avouris}. It is interesting that
graphene with a zigzag edge has the half-filled flat band at the zero-energy
level and show edge ferromagnetism\cite{Fujita}. The half-filled zero-energy
states emerge also in all zigzag nanoribbons and hence they are metallic\cite%
{Fujita,EzawaPRB}. However, realistic nanoribbons have finite length. It is
important to investigate the finite-length effects on the electronic
properties of nanoribbons.

\begin{figure}[t]
\centerline{\includegraphics[width=0.4\textwidth]{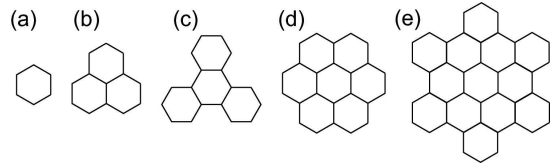}}
\caption{Basic configurations of typical graphene nanodisks. (a) Benzene.
(b) Trigonal zigzag nanodisk (phenalene). (c) Trigonal armchair nanodisk
(triphenylene). (d) Hexagonal zigzag nanodisk (coronene). (e) Hexagonal
armchair nanodisk (hexa benzocoronene)\protect\cite{Rader}. }
\label{FigNanodisk}
\end{figure}

\begin{figure}[t]
\begin{center}
\includegraphics[width=0.3\textwidth]{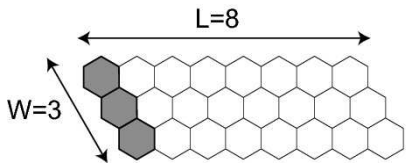}
\end{center}
\caption{Geometric configuration of zigzag nanoribbons with width $W$ and
length $L$. Here we show the example of the $(W,L)=(3,8)$ nanoribbon. The
basic chain is $W$ connected benzene depicted in gray. Short nanoribbons may
be regarded as parallelogrammic nanodisks.}
\label{FRibbon}
\end{figure}

Another basic element of graphene derivatives is a graphene nanodisk\cite%
{EzawaPhysica}. It is a nanometer-scale disk-like material which has a
closed edge. A graphene nanodisk can be constructed by connecting several
benzenes. There are many type of nanodisks, where typical examples are
displayed in Fig.\ref{FigNanodisk}. Finite-length nanoribbons may be
regarded as nanodisks provided that their length is short [Fig.\ref{FRibbon}%
]. Some of nanodisks have already been manufactured by soft-landing mass
spectrometry\cite{Rader}.

In this paper we analyze the electric properties of nanodisks as well as
finite-length nanoribbons. Since all zigzag nanoribbons are metallic, as we
have mentioned, we expect all zigzag graphene derivatives are also metallic.
On the contrary, the emergence of zero-energy states is quite rare. We show
that there are no zero-energy states in finite-length zigzag nanoribbons. We
also investigate a class of trigonal and hexagonal nanodisks possessing
zigzag or armchair edges, among which we have found zero-energy states only
in trigonal zigzag nanodisks.

Trigonal zigzag nanodisks are prominent in their electronic properties
because there exist half-filled zero-energy states and they are metallic.
Indeed, we can engineer nanodisks equipped with an arbitrary number of
degenerate zero-energy states. Furthermore, spins are argued to make a
ferromagnetic coupling. A remarkable property is that the relaxation time is
quite large in spite of its small size.

This paper is organized as follows. In Section \ref{EnergRibbo} and \ref%
{EnergDisk}, based on the tight-binding Hamiltonian, we calculate the energy
spectra of finite-length zigzag nanoribbons and of a wide class of graphene
nanodisks, respectively. In Section \ref{EnergDisk}, we also carry out a
systematic investigation of the zero energy states in trigonal zigzag
nanodisks. In Section \ref{Wave}, we analyze the wave functions of these
zero-energy states to examine how they are localized at the edges. In
Section \ref{Magnet}, we study the spin-relaxation time of nanodisks.
Section \ref{Discuss} is devoted to discussions.

\section{Energy spectrum of Finite-Length Nanoribbons}

\label{EnergRibbo}

We calculate the energy spectra of graphene derivatives based on the
nearest-neighbor tight-binding model, which has been successfully applied to
the studies of carbon nanotubes\cite{Saito} and nanoribbons\cite{EzawaPRB}.
The Hamiltonian is defined by%
\begin{equation}
H=\sum_{i}\varepsilon _{i}c_{i}^{\dagger }c_{i}+\sum_{\left\langle
i,j\right\rangle }t_{ij}c_{i}^{\dagger }c_{j},  \label{HamilTB}
\end{equation}%
where $\varepsilon _{i}$ is the site energy, $t_{ij}$ is the transfer
energy, and $c_{i}^{\dagger }$ is the creation operator of the $\pi $
electron at the site $i$. The summation is taken over all nearest
neighboring sites $\left\langle i,j\right\rangle $. Owing to their
homogeneous geometrical configuration, we may take constant values for these
energies, $\varepsilon _{i}=\varepsilon _{\text{F}}$ and $t_{ij}=t$. Then,
the diagonal term in (\ref{HamilTB}) yields just a constant, $\varepsilon _{%
\text{F}}N_{\text{C}}$, where $N_{\text{C}}$ is the number of carbon atoms
in the system. The Hamiltonian (\ref{HamilTB}) yields the Dirac electrons
for graphene\cite{GraphExA,GraphExB,GraphExC}. There exists one electron per
one carbon and the band-filling factor is 1/2. It is customary to choose the
zero-energy level of the tight-binding Hamiltonian (\ref{HamilTB}) at this
point so that the energy spectrum is symmetric between the positive and
negative energy states. Therefore, the system is metallic provided that
there exists zero-energy states in the spectrum.

In this section we investigate finite-length nanoribbons to know if there
are zero-energy states. A classification of infinite-length nanoribbons is
given in a previous work\cite{EzawaPRB}. Here we concentrate on
finite-length zigzag nanoribbons. We classify them as follows (Fig.\ref%
{FRibbon}). First we take a basic chain of $W$ connected carbon hexagons, as
depicted in dark gray. Second we translate this chain. Repeating this
translation $L$ times we construct a nanoribbon indexed by a set of two
integers $\left( W,L\right) $. In what follows we analyze a class of
finite-length nanoribbons generated in this way. Parameters $W$ and $L$
specify the width and the length of nanoribbons, respectively. The
infinite-length nanoribbons are obtained by letting $L\rightarrow \infty $.
The finite-length nanoribbons are regarded as parallelogrammic nanodisks
when $L\approx W$.

\begin{figure}[t]
\begin{center}
\includegraphics[width=0.47\textwidth]{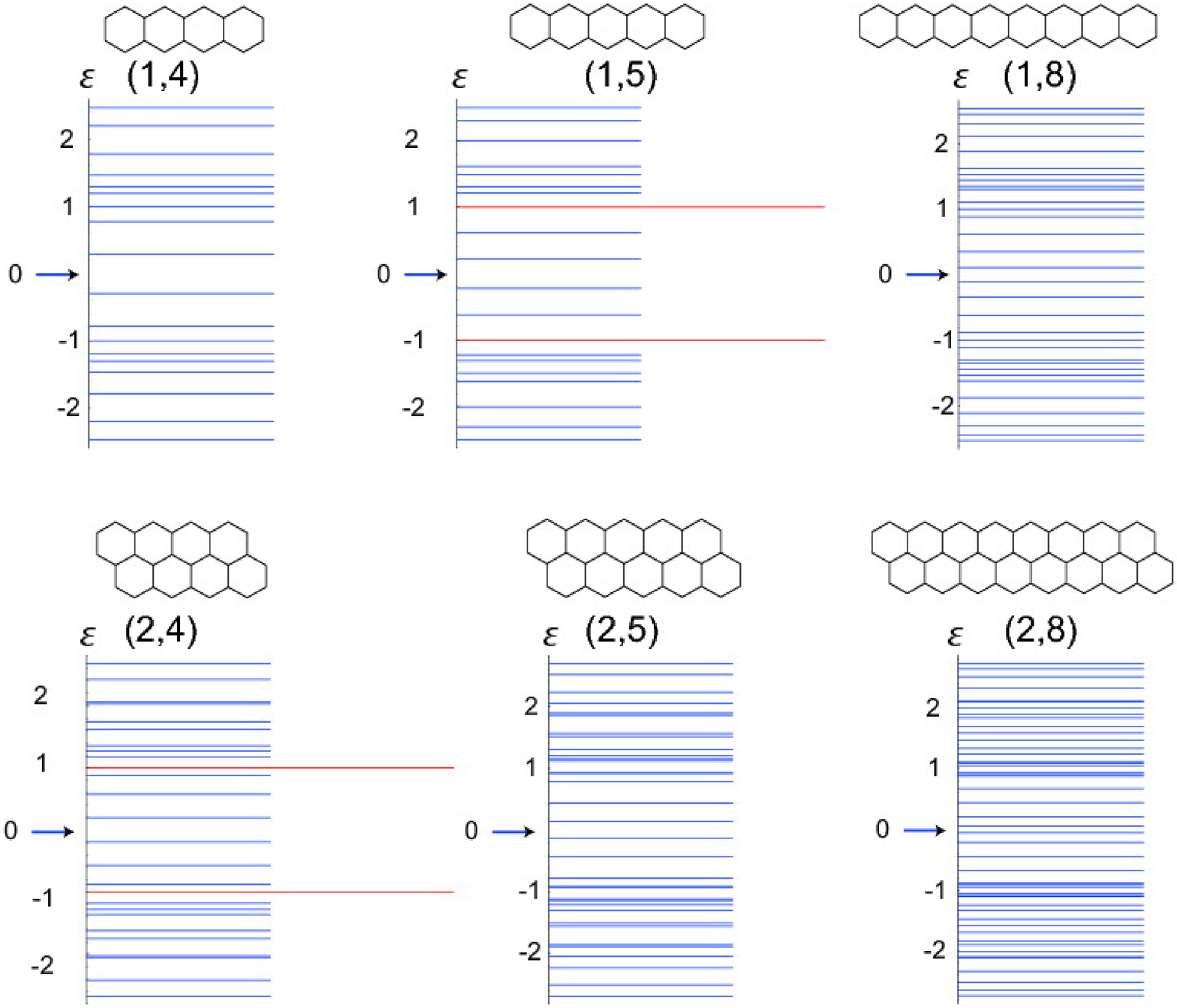}
\end{center}
\caption{(Color online) Density of states of finite-length nanoribbons. The
vertical axes is the energy $\protect\varepsilon $ in units of $t=3$eV, and
the horizontal axes is the degeneracy. There exist no zero-energy states.}
\label{FDOS}
\end{figure}

In analyzing a nanoribbon containing $N_{\text{C}}$ carbon atoms, the
Hamiltonian (\ref{HamilTB}) is reduced to an $N_{\text{C}}\times N_{\text{C}%
} $ matrix. It is possible to diagonalize the Hamiltonian exactly to
determine the energy spectrum $E_{i}$ together with its degeneracy $g_{i}$
for each finite-length nanoribbon. The density of state is given by 
\begin{equation}
D\left( \varepsilon \right) =\sum_{i=1}^{N_{\text{C}}}g_{i}\delta \left(
\varepsilon -E_{i}\right) .  \label{DOS}
\end{equation}%
It can be shown that the determinant associated with the Hamiltonian (\ref%
{HamilTB})\ has a factor such that%
\begin{equation}
\det \left[ \varepsilon I-H\left( N_{\text{C}}\right) \right] \propto
(\varepsilon -t)^{a\left( W,L\right) }(\varepsilon +t)^{a\left( W,L\right) },
\end{equation}%
implying the $a\left( W,L\right) $-fold degeneracy of the states with the
energy $\varepsilon =\pm t$, where\beginABC%
\begin{eqnarray}
a\left( 1,L\right) &=&2,1,2,1,2,1,2,1,2,1,...., \\
a\left( 2,L\right) &=&1,1,0,2,0,1,1,0,2,0,...., \\
a\left( 3,L\right) &=&2,0,2,0,2,0,2,0,2,0,...., \\
a\left( 4,L\right) &=&1,2,0,3,0,2,1,1,2,0,.....
\end{eqnarray}%
\endABC{}We have displayed the full spectra for some examples of
finite-length nanoribbons by taking $t=3$eV in Fig.\ref{FDOS}.

\begin{figure}[t]
\begin{center}
\includegraphics[width=0.35\textwidth]{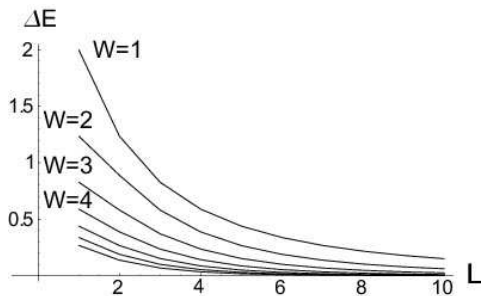}
\end{center}
\caption{Band gap of zigzag nanoribbons as a function of length $L$. The
horizontal axes is the length $L$ and the vertical axes is the energy gap $%
\Delta E$ in units of $t=3$eV. Each curve is for width $W=1$ to $7$ from top
to bottom.}
\label{FGap}
\end{figure}

One of our main results is that there are no zero-energy states in
finite-length nanoribbons. However, the band gap decreases inversely to the
length, and zero-energy states emerge as $L\rightarrow \infty $, as shown in
Fig.\ref{FGap}. This is consistent with the fact that infinite-length
nanoribbons have the flat band made of degenerated zero-energy states\cite%
{Fujita,EzawaPRB}. Hence, a sufficiently long nanoribbon can be regarded
practically as a metal.

\section{Energy spectrum of Nanodisks}

\label{EnergDisk}

\begin{figure}[t]
\centerline{\includegraphics[width=0.5\textwidth]{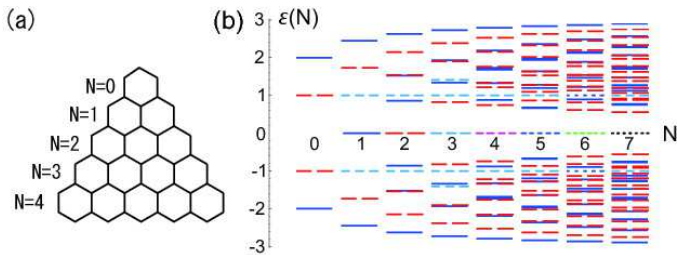}}
\caption{(Color online) (a) Geometric configuration of trigonal zigzag
nanodisks. It is convenient to introduce the size parameter $N$ in this way.
The $0$-trigonal nanodisk consists of a single Benzene, and so on. The
number of carbon atoms are related as $N_{\text{C}}=N^{2}+6N+6$. See
explicit examples given in Fig.\protect\ref{FigNanoStair}. (b) Density of
states of the $N$-trigonal nanodisk for $N=0,1,2,\cdots ,7$. The horizontal
axis is the size $N$ and the vertical axis is the energy $\protect%
\varepsilon (N)$ in units of $t=3$eV. Dots on colored bar indicate the
degeneracy of energy levels.}
\label{FigDotSamp}
\end{figure}

We next derive the energy spectrum of each nanodisk [Fig.\ref{FigNanodisk}]
by diagonalizing the Hamiltonian (\ref{HamilTB}). As an example we display
the density of state (\ref{DOS}) of trigonal zigzag nanodisks in Fig.\ref%
{FigDotSamp}. We have classified them by the size parameter $N$ as defined
in Fig.\ref{FigDotSamp}(a). The number of carbons are given by $N_{\text{C}%
}=N^{2}+6N+6$.

\begin{figure}[t]
\centerline{\includegraphics[width=0.4\textwidth]{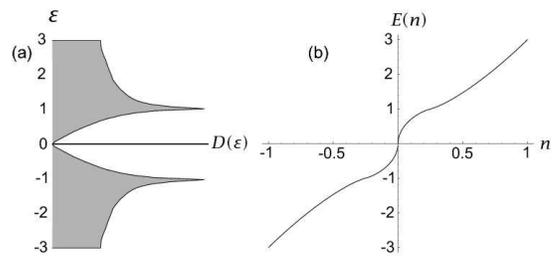}}
\caption{(a) Density of state $D(\protect\varepsilon )$ as a function of the
energy $\protect\varepsilon $ in units of $t=3$eV for graphene. (b) The
energy $E(n)$ as a function of the doped electron number $n$. These two
functions share the same information of the energy spectrum.}
\label{FigGrapDOS}
\end{figure}

In order to reveal a global structure, it is convenient to introduce the
doped electron number at a given energy $E$, which we normalize as 
\begin{equation}
n\left( E\right) =\frac{1}{N_{\text{C}}}\int_{0}^{E}D\left( \varepsilon
\right) d\varepsilon ,
\end{equation}%
with $|n(E)|\leq 1$. We then make the inversion of this formula to derive $E$
as a function of $n$. See Fig.\ref{FigGrapDOS} for their correspondence in
graphene, where $E\left( n\right) $ is found to be a prolonged S-shaped
curve.

\begin{figure}[t]
\centerline{\includegraphics[width=0.4\textwidth]{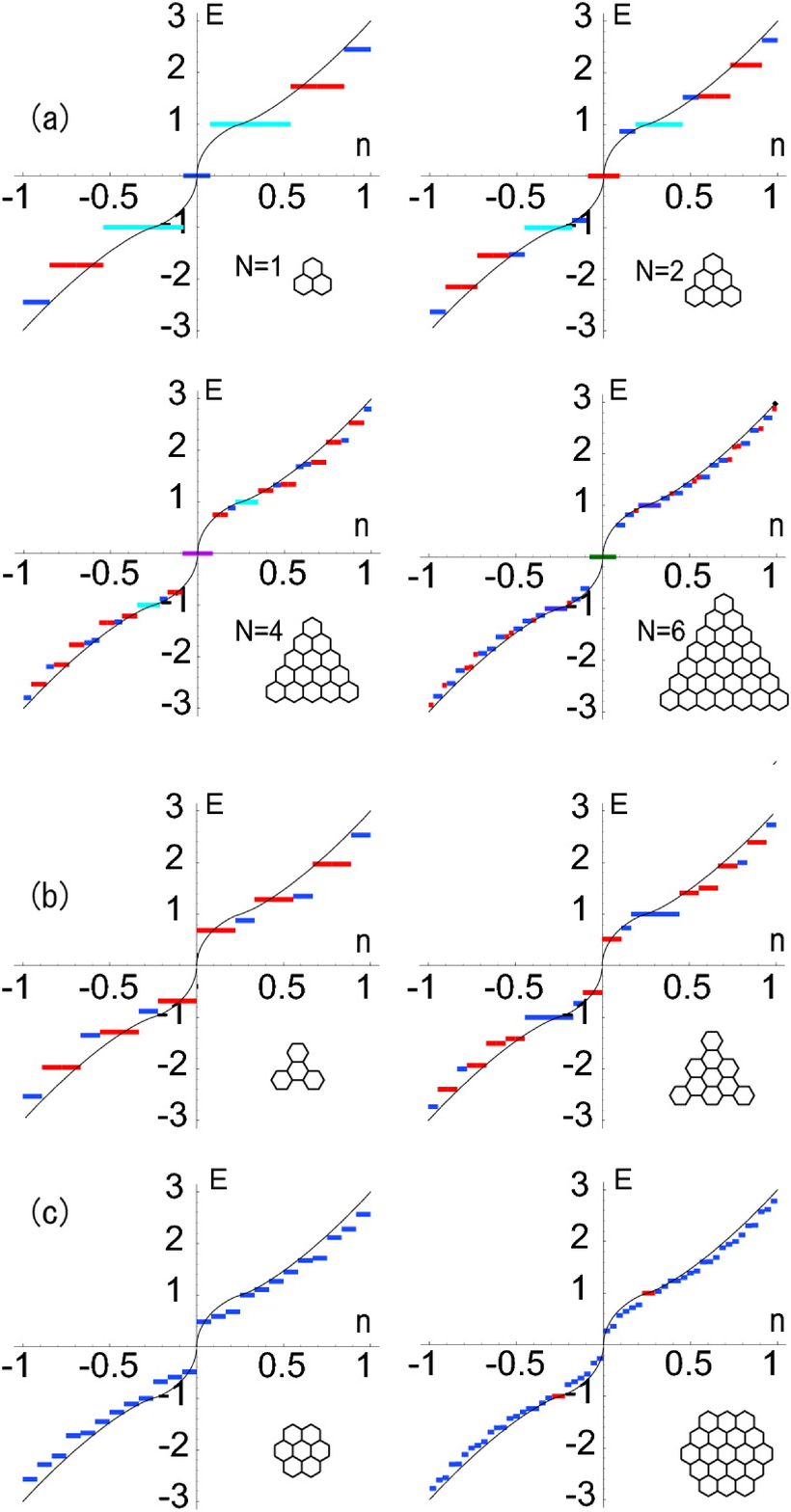}}
\caption{(Color online) Energy spectrum $E(n)$ as a function of the doped
electron number $n $ for (a) trigonal zigzag nanodisks, (b) trigonal
armchair nanodisks and (c) hexagonal zigzag nanodisks, with various sizes.
The horizontal axis is the number of doped electrons $n$, and the vertical
axis is the energy $E$ in units of $t=3$eV. The S-shaped solid curve is that
of graphene. The energy density diagrams of nanodisks approach that of
graphene for large size. There are degenerate zero-energy states in all
trigonal nanodisks, and they are metallic. There are no zero-energy states
in all zigzag armchair nanodisks and all hexagonal nanodisks, and they are
semiconducting. }
\label{FigNanoStair}
\end{figure}

Diagonalizing the Hamiltonian (\ref{HamilTB}) explicitly, we have
constructed and displayed $E(n)$ for several nanodisks with trigonal zigzag
shape in Fig.\ref{FigNanoStair}(a), trigonal armchair shape in Fig.\ref%
{FigNanoStair}(b) and hexagonal zigzag shape in Fig.\ref{FigNanoStair}(c).
Each diagram consists of step-like segments reflecting the $\delta $%
-function type density of states (\ref{DOS}). The length of each step
represents the degeneracy of the energy level in units of $N_{\text{C}}$. It
is remarkable that there exist zero-energy states only in trigonal zigzag
nanodisks. We have also checked explicitly the absence of the zero-energy
state in a series of nanodisks with hexagonal armchair type [Fig.\ref%
{FigNanodisk}(e)].

In each figure we have also displayed the prolonged S-shaped curve of
graphene, which the $E\left( n\right) $ of nanodisk approaches in the large
size limit ($N_{\text{C}}\rightarrow \infty $). The prolonged S-shaped curve
is universal regardless of the nanodisk's shape.

We investigate trigonal zigzag nanodisks more in details since they have
zero-energy states. It can be shown that the determinant associated with the
Hamiltonian (\ref{HamilTB})\ has a factor such that%
\begin{equation}
\det \left[ \varepsilon I-H\left( N_{\text{C}}\right) \right] \propto
\varepsilon ^{N}(\varepsilon -t)^{a\left( N\right) }(\varepsilon
+t)^{a\left( N\right) },
\end{equation}%
implying the $N$-fold degeneracy of the zero-energy states and the $a\left(
N\right) $-fold degeneracy of the states with the energy $\varepsilon =\pm t$%
, where%
\begin{equation}
a\left( N\right) =3,3,3,3,3,5,3,5,3,7,3,7,3,9,3,9,3,....,
\end{equation}%
for $N=1,2,3,\cdots $.

Since there exist half-filled zero-energy states for $N\geq 1$, these
nanodisks are metallic. The gap energy between the first excitation state
and the ground state decreases as the size becomes larger. However, it is
remarkable that the gap energy is quite large and is of the order of $3$eV
even in the nanodisk with size $N=7$ [Fig.\ref{FigDotSamp}(b)]. This is much
higher than room temperature. Hence, the low-energy physics near the Fermi
energy $\varepsilon =0$ can well be described only by taking the zero-energy
states into account.

\begin{figure}[t]
\centerline{\includegraphics[width=0.4\textwidth]{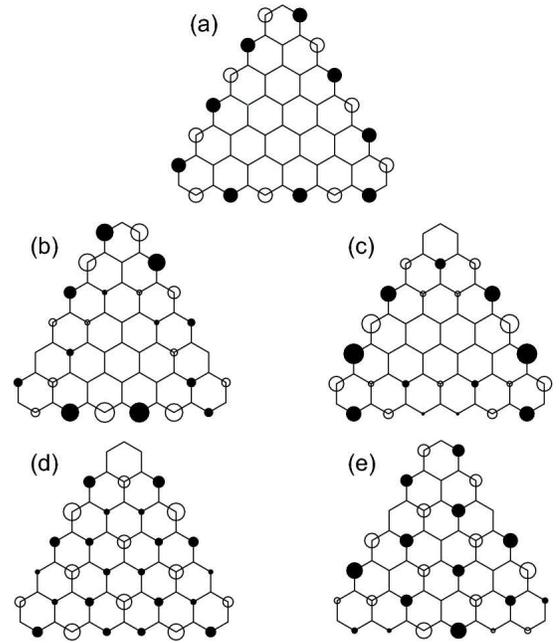}}
\caption{The zero-energy states of the trigonal nanodisk with size $N=5$.
There are $5$ degenerate states, (a) $\sim $ (e). Electrons are localized on
edges in the state (a). When the site energy $\protect\varepsilon _{i}$ is
decreased at edges equally, the degeneracy is partially resolved, as
illustrated in Fig.\protect\ref{FigZEdge}(a). The state (a) has the lowest
energy and nondegenerate; the states (b) and (c) are degenerate; the states
(d) and (e) are degenerate and have the highest energy. When the site energy 
$\protect\varepsilon _{i}$ is decreased further on the bottom edge, all the
degeneracy among these 5 states is resolved, as illustrated in Fig.\protect
\ref{FigZEdge}(b).}
\label{FigWave}
\end{figure}

\section{Wave functions of trigonal nanodisks}

\label{Wave}

A wave function is represented as%
\begin{equation}
\varphi (\boldsymbol{x})=\sum_{i}\omega _{i}\varphi _{i}(\boldsymbol{x}),
\end{equation}%
where $\varphi _{i}(\boldsymbol{x})$ is the Wannier function localized at
the lattice point $i$. The operator $c_{i}$ in the Hamiltonian (\ref{HamilTB}%
) annihilates an electron in the state described by the Wannier function $%
\varphi _{i}(\boldsymbol{x})$. We are able to calculate the amplitude $%
\omega _{i}$ for zero-energy states in the trigonal zigzag nanodisk. All of
them are found to be real. As an example we show them with size $N=5$ in Fig.%
\ref{FigWave}, where the solid (open) circles denote the amplitude $\omega
_{i}$ are positive (negative). The amplitude is proportional to the radius
of circle. It is intriguing that one of the wave functions is entirely
localized on edge sites for nanodisks with $N=$odd, as in Fig.\ref{FigWave}%
(a). There are no such wave functions for nanodisks with $N=$even.

\begin{figure}[t]
\centerline{\includegraphics[width=0.5\textwidth]{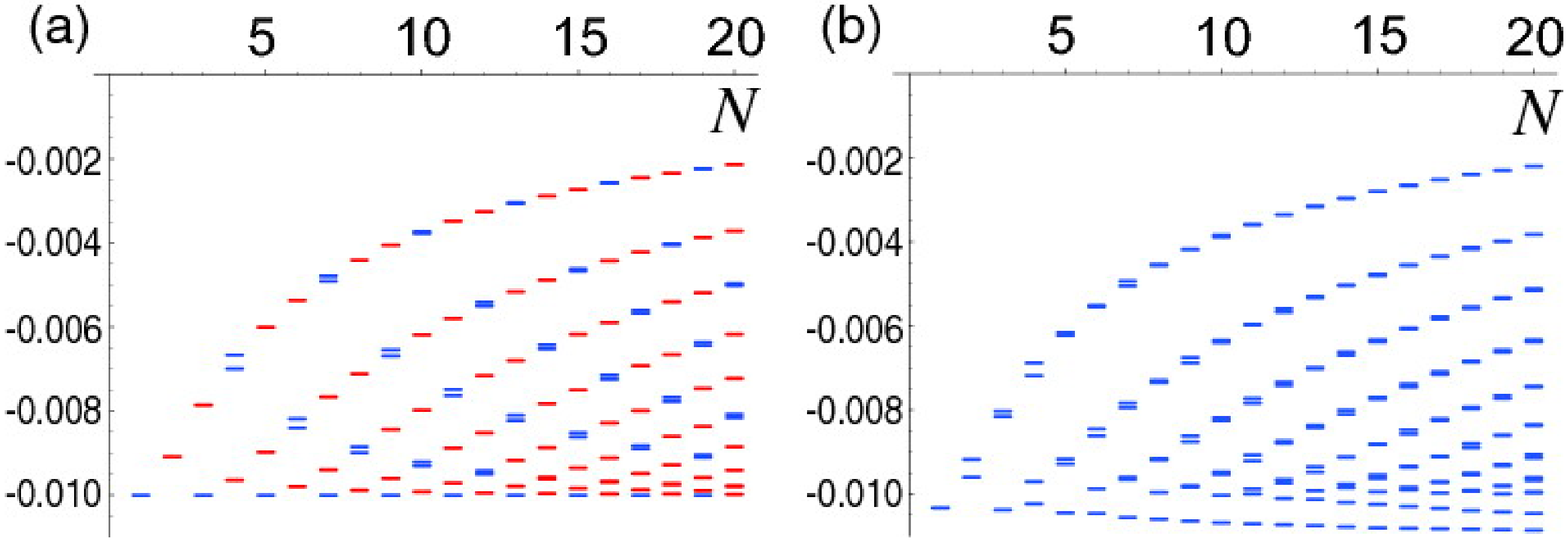}}
\caption{(Color online) Zero-energy states split into several nonzero-energy
states by edge modifications. \ The horizontal axis is the size $N$ and the
vertical axis is the energy in units of $t=3$eV. (a) We take $\protect%
\varepsilon _{i}=\protect\varepsilon -\Delta \protect\varepsilon $ for all
edge carbons with $\Delta \protect\varepsilon =0.03$eV. Nonzero-energy
states are nondegenerate (blue) or 2-fold degenerate (red). (b) We take $%
\protect\varepsilon _{i}=\protect\varepsilon -\Delta \protect\varepsilon %
-\Delta \protect\varepsilon ^{\prime }$, $\Delta \protect\varepsilon %
^{\prime }=0.003$eV for edge carbons on only one of the three edges and $%
\protect\varepsilon _{i}=\protect\varepsilon -\Delta \protect\varepsilon $
for those on the other two edges. All states become nondegenerate.}
\label{FigZEdge}
\end{figure}

The most important property is that all wave functions are nonvanishing on
edge sites. In order to demonstrate this, we have investigated how the
zero-energy states are modified by changing the site-energy $\varepsilon
_{i} $ in the Hamiltonian (\ref{HamilTB}) only for edge carbons. Edge
carbons are those surrounded by two carbon atoms and one hydrogen atom,
while bulk carbons are those surrounded by three carbon atoms\cite{EzawaPRB}%
. If a wave function vanishes on edges, the zero-energy state must remain as
it is. First, we take $\varepsilon _{i}=\varepsilon -\Delta \varepsilon $
for all edge carbons. We show how the zero-energy states split by taking $%
\Delta \varepsilon =0.01t=0.03$eV in Fig.\ref{FigZEdge}(a) for $N=1,2,\cdots
,20$. It is seen that all zero-energy states acquire negative energy and
that they become one-fold or two-fold degenerate. The energy decrease is
larger when the wave function is localized more on edges. When we decrease
the site-energy more, $\varepsilon _{i}=\varepsilon -\Delta \varepsilon
-\Delta \varepsilon ^{\prime }$, $\Delta \varepsilon ^{\prime }=0.001t=0.003$%
eV for edge carbons on only one of the three edges, all the degeneracy is
found to be resolved as in Fig.\ref{FigZEdge}(b).

\section{Magnetic properties of trigonal nanodisks}

\label{Magnet}

We proceed to show that the degenerate ground states lead to a
ferromagnetism as in graphene nanoribbon. This is because the Coulomb
exchange interaction or the Hund's rule coupling drives all spins to
polarize into a single direction. We are most interested how large is the
relaxation time for small nanodisks. (Strictly speaking, ferromagnetism can
occur only in an infinitely large system, but we may use the terminology for
a finite system if the relaxation time is large enough.)

The effective Hamiltonian for the Coulomb exchange interaction is given by
the Heisenberg model. Here, for the sake of simplicity, we use the Ising
model,%
\begin{equation}
H=-\sum_{i\neq j}^{N}J_{ij}\sigma _{i}\sigma _{j},  \label{Ising}
\end{equation}%
since the Heisenberg model presents essentially the same result on the
relaxation time as we shall argue later. In the effective Hamiltonian, $%
\sigma _{i}$ is the spin operators of electrons in the $i$th zero-energy
state, $\sigma _{i}=\pm 1$, and the summation is taken over all electron
pairs. An important point is that the exchange interaction strength $J_{ij}$
must be nonzero for all electron pairs because their wave functions are
nonvanishing on edges. This is in a sharp contrast to the Hamiltonian for a
nanomagnet, where $J_{ij}$ can be regarded nonvanishing only for neighboring
electron pairs since the index $i$ represents the site in the real space.
This makes a clear difference in the relaxation time as we shall soon see.

The spin dynamics is well described by the master equation, $dP\left(
t\right) /dt=\mathcal{L}P\left( t\right) $, where the symbol $P$\ denotes
the probability distribution function specifying the spin configuration, and 
$\mathcal{L}$ is the Liouville operator associated with the Hamiltonian.
When the expansion $P\left( 0\right) =\sum_{\lambda }c_{\lambda }P_{\lambda
} $ holds at the initial state at time $t=0$, the state of the system at any
later time is given by%
\begin{equation}
P\left( t\right) =e^{\mathcal{L}t}P\left( 0\right) =\sum_{\lambda
}c_{\lambda }e^{-\lambda t}P_{\lambda },
\end{equation}%
where $\lambda $ is obtained by solving the eigenvalue equation, $\mathcal{L}%
P_{\lambda }=-\lambda P_{\lambda }$. The relaxation rate of the system is
equal to the minimum eigenvalue $\lambda _{\text{min}}$ of the Liouville
operator. Thus the relaxation time is given by $\tau =1/\lambda _{\text{min}%
} $.

\begin{figure}[t]
\centerline{\includegraphics[width=0.5\textwidth]{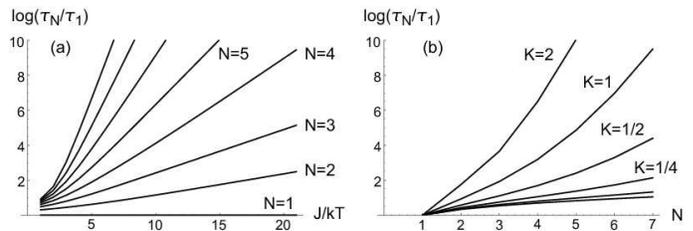}}
\caption{Relaxation time of various graphene nanodisks. (a) The relaxation
time of nanodisks with size $N=1,2,\cdots ,8$ from bottom to top. The
horizontal axis is the coupling constant $J/kT$ and the vertical axis is the
relaxation time in the form of log$_{10}$[$\protect\tau _{N}/\protect\tau %
_{1}$]. (b) The relaxation time of nanodisks with interaction strength $%
K\equiv J/kT=1/16,1/8,1/4,1/2,1,2$ from bottom to top. The horizontal axis
is the size of nanodisks $N$. }
\label{FigRelTime}
\end{figure}

To get a concrete idea, since all $J_{ij}$ are nonvanishing, we first make
an estimation by making an approximation $J_{ij}=J$. We are able to
diagonalize the Ising Hamiltonian (\ref{Ising}) explicitly. The eigenvalues
are given by%
\begin{equation}
E_{n}=-\frac{J}{2}\left[ \left( N-2n\right) ^{2}-N\right] ,
\label{IsingEnerg}
\end{equation}%
where $n=0,1,2,\cdots ,N$ is the energy level index. We have then carried
out an exact diagonalization of the eigenvalue problem of the Liouville
operator $\mathcal{L}$ and determined the eigenvalue $\lambda _{\text{min}}$%
. We show the relaxation time as a function of the coupling strength $J/kT$
for various size $N$\ in Fig.\ref{FigRelTime}(a) and as a function of size
for various coupling strength in Fig.\ref{FigRelTime}(b).

For the noninteracting case, $J/kT=0$, it is easy to see that the relaxation
time $\tau _{N}$ of the $N$-trigonal zigzag nanodisk is given by%
\begin{equation*}
\tau _{N}=N\tau _{1}.
\end{equation*}%
On the other hand, the relaxation rate $\lambda _{\text{min}}$ is given by
the Arrhenius-type formula for strong coupling limit or in low temperature
limit, $J/kT\gg 1$. Hence the relaxation time is given by%
\begin{equation}
\tau _{N}=\exp \left[ \frac{\Delta E}{kT}\right] \tau _{1},
\label{RelaxTime}
\end{equation}%
where%
\begin{equation}
{}%
\begin{array}{ll}
\displaystyle\Delta E=JN^{2}/2 & \quad \text{for }N=\text{even,} \\[4mm] 
\displaystyle\Delta E=J(N^{2}-1)/2 & \quad \text{for }N=\text{odd}%
\end{array}%
\end{equation}%
is the energy difference between the highest energy state and the ground
state. It is observed in Fig.\ref{FigRelTime} that the relaxation time is
given by the asymptotic formula (\ref{RelaxTime}) already for $J\gtrsim kT$.

The relaxation rate is given by the Arrhenius-type formula also in the
generic model (\ref{Ising}) with $J_{ij}\neq J$, where $\Delta E$ is the
energy difference between the highest energy state and the ground state.
Then, the relaxation time is given by (\ref{RelaxTime}) by replacing $J$
with 
\begin{equation}
J_{\text{eff}}\simeq \frac{1}{N(N-1)}\sum_{i\neq j}^{N}J_{ij},
\label{EffecJ}
\end{equation}%
which is of the order of a typical $J_{ij}$. Furthermore, we would obtain
the same result for the Heisenberg model even with $J_{ij}\neq J$.

Note that in an ordinary nanomagnet composed of $N$ spins, the relaxation
time is given by (\ref{RelaxTime}) with%
\begin{equation}
\Delta E=JNz,
\end{equation}%
where $z$ is the number of the nearest neighboring spins. It is remarkable
that the size dependence of the relaxation time is $\propto N^{2}$ for
trigonal zigzag nanodisks, though it is $\propto N$ for normal nanomagnets.
This is because any one spin couples with all other spins in the zero-energy
state.

\section{Discussions}

\label{Discuss}

Graphene derivatives become metallic when they have half-filled zero-energy
states. We have explored the energy spectra in a wide class of nanodisks as
well as finite-length nanoribbons. It is surprising that the emergence of
zero-energy states is quite rare. There exist no zero-energy states in
finite-length zigzag nanoribbons. However, the band gap decreases inversely
to the length, and zero-energy states emerge as $L\rightarrow \infty $.
Hence, a sufficiently long nanoribbon can be regarded practically as a metal.

Among a wide class of nanodisks we have studied, trigonal zigzag nanodisks
are distinguished\ for their electronic properties since they exhibit
metallic ferromagnetism due to their half-filled degenerate zero-energy
states. The degeneracy is controllable arbitrarily by changing the size of
nanodisks. We have estimated the relaxation time, which has been argued to
be proportional to $\exp [J_{\text{eff}}N^{2}/2kT]$ when the size is $N$.
Though the numerical estimation of the effective spin stiffness $J_{\text{eff%
}}$ is yet to be done, it is of the order of the Coulomb energy since its
origin is the exchange interaction or the Hund's coupling rule. We conclude
that the relaxation time is quite large for its size at low temperature $%
T\lesssim J_{\text{eff}}/2k$. Hence, for instance, it could be used as a
memory device. By connecting nanodisks with nanoribbons, we can design
electronic circuits. These devices would be obtained by etching a single
graphene. Alternatively, nanodisks may be connected with leads by making
tunneling junctions. We would like to make an analysis of the nanodisk and
lead system together with related phenomena in future works.

I am very much grateful to Professors Y. Hirayama and K. Hashimoto for many
fruitful discussions on the subject. The work was in part supported by
Grants-in-Aid for Scientific Research from Ministry of Education, Science,
Sports and Culture (Nos.070500000466).

\end{document}